\documentclass[fleqn,twoside]{article}
\usepackage{espcrc2}

\usepackage{graphicx}
\usepackage[figuresright]{rotating}
\input{epsf}
\newcommand{\beq}{\begin{equation}}
\newcommand{\eeq}{\end{equation}}
\newcommand{\bea}{\begin{eqnarray}}
\newcommand{\eea}{\end{eqnarray}}
\newcommand{\bmp}{\begin{minipage}}
\newcommand{\emp}{\end{minipage}}

\newcommand{\tr}{{\rm tr}}

\newcommand{\AmS}{{\protect\the\textfont2
  A\kern-.1667em\lower.5ex\hbox{M}\kern-.125emS}}

\title{From short to long scales in the QCD vacuum}

\author{E.T. Tomboulis\address{Department of Physics, UCLA, Los Angeles, 
CA 90095-1547, USA}%
        \thanks{Research partially supported by NSF-PHY-0309362.}}
       
\begin{document}

\begin{abstract}
We study approximate decimations in SU(N) LGT that connect the 
short to long distance regimes, and provide both upper 
and lower bounds on the exact partition function. 
This leads to a representation of the exact partition function in  terms of 
successive decimations. The implications for a derivation of confinement  
from first principles are discussed.
\vspace{1pc}
\end{abstract}

\maketitle

A very large body of numerical and analytical 
work has been done by the lattice community in 
recent years on the types of configurations 
responsible for confinement. A great deal of 
information concerning the confinement mechanism   
has been obtained from these investigations (for recent review, 
see \cite{Gr}). 
However, the goal of a direct derivation of confinement from first principles 
has remained elusive for the last thirty years. 

The origin of the difficulty is clear. One is faced with 
a multi-scale problem: passage from the short-distance   
weakly coupled, ordered regime to the long distance strongly 
coupled, disordered, confining regime. 
Such variable multi-scale behavior can only be addressed by some 
nonperturbative block-spinning or decimation procedure. Exact 
decimation schemes appear analytically hopeless, and numerically very 
difficult. Here we will consider simple `bond moving' decimations 
and show that they can provide bounds on the exact theory,  
allowing statements about its behavior and  
the question of an actual derivation of confinement in LGT.

Starting with some plaquette action, e.g the Wilson action 
$A_p(U) ={\beta\over N}\;{\rm Re}\,\tr U_p$, at lattice spacing $a$, 
we consider the character expansion:
\beq 
F(U, a) =e^{A_p(U)} 
   = \sum_j\;F_j(\beta,a)\,d_j\,\chi_j(U) 
\eeq
with Fourier coefficients: 
\[ F_j = \int\,dU\;F(U,a)\,{1\over d_j}\,\chi_j(U)\; .\]
E.g, for SU(2), $j=0, {1\over 2}, 1, {3\over 2}, \ldots$, $d_j=(2j+1)$.
In terms of normalized coefficients:  
\bea
F(U, a) &=&  F_0\,\Big[\, 1 + \sum_{j\not= 0} c_j(\beta)\,d_j\,\chi_j(U)\,
 \Big] \nonumber \\
    & \equiv & F_0\;f(U,a)   
\eea 
For a reflection positive action one has: 
\beq
F_j \geq 0\qquad \mbox{hence}\quad 1\geq c_j\geq 0 \qquad\quad 
\mbox{all}\quad j \;.
\eeq
The partition function (PF) on lattice $\Lambda$ is then  
\[ Z_\Lambda(\beta) 
                    =F_0^{|\Lambda|}\; \int dU_\Lambda\;\prod_p\,f_p(U,a)
\]

We now consider RG decimation 
transformations $a \to \lambda a$.  
Simple approximate transformations of the `bond moving' type 
are implemented by  
`weakening', i.e. decreasing the $c_j$'s  of interior plaquettes, 
and `strengthening', i.e. increasing $c_j$'s  of boundary plaquettes 
in every decimation cell of side length $\lambda$. 
The simplest scheme \cite{MK}, which is  
adopted in the following,  implements  
complete removal, $c_j=0$, of interior plaquettes.

Under successive decimations 
\bea 
& & a \to \lambda a \to \lambda^2 a \to \cdots \to \lambda^n a \nonumber \\
 & & \Lambda \to \Lambda^{(1)} \to \Lambda^{(2)} \to \cdots \to 
\Lambda^{(n)} \nonumber  
\eea
the RG transformation rule is then:
\beq
f(U,n-1)\to f(U,n) = \Big[ 1 + \sum_{j\not= 0} c_j(n)\,d_j
\,\chi_j(U) \Big] \label{RG1}
\eeq
with:  
\beq
c_j(n) = F_j(n)/F_0(n)\, , \quad F_j(n)=
\Big[\hat{F}_j(n)\Big]^{\lambda^2} , \label{RG2}
\eeq   
\beq
\hat{F}_j(n)=  \int\,dU\;\Big[\,f(U,n-1)\,\Big]^\nu\,{1\over d_j}\,\chi_j(U) 
\; . \label{RG3}
\eeq
The parameter $\nu$ controls by how much the remaining plaquettes 
have been strengthened to compensate for the removed plaquettes. 
The resulting PF after $n$ decimation steps is:  
 \[ Z_\Lambda(\beta, n) 
                    =\prod_{m=0}^n F_0(m)^{|\Lambda|/\lambda^{md}}\; 
\int dU_{\Lambda^{(n)}}\;\prod_p\,f_p(U,n)
\]

It is important to note that after each decimation step 
the resulting action retains the 
original {\it one-plaquette form} but will, in general,  
contain all representations:
\[ A_p(U,n)= \sum_j \; \tilde{\beta}_j(\beta)\,\chi_j(U) \;.\]
Also, both {\it positive and negative} effective couplings 
$\tilde{\beta}_j$ will occur. 
This is the case even after a single decimation step $a\to \lambda a$ 
starting with the Wilson action. 

However, for {\it integer} $\nu$, 
the property $ F_0(n)\geq 0$, $c_j(n)\geq 0$ and hence 
reflection positivity are maintained at each decimation step.

The subsequent development hinges on the following two basic 
statements that can now be proven: 

(I) With $\nu=\lambda^{d-2}$:  
\[ Z_\Lambda(\beta, n) \leq Z_\Lambda(\beta, n+1) \;.\] 

(II) With $\nu=1$: 
\[  Z_\Lambda(\beta, n+1) \leq Z_\Lambda(\beta, n) \; .\] 
In fact, in both (I), (II) one has strict inequality. 

(II) says that decimating plaquettes while leaving the couplings of the 
remaining plaquettes unaffected results in a lower bound on the P.F. 
Reflection positivity (positivity of Fourier coefficients) is crucial 
for this to hold. 

(I) says that modifying the couplings of the remaining plaquettes after 
decimation by taking $\nu=\lambda^{d-2}$ (standard MK choice \cite{MK}) 
results into overcompensation (upper bound on the P.F.).

Consider now the, say, $(n-1)$-th decimation step  
with Fourier coefficients $c_j(n-1)$, which we relabel   
$c_j(n-1)=\tilde{c}_j(n-1)$.  
Given these $\tilde{c}_j(n-1)$, we proceed to compute the coefficients 
$F_0(n)$, $c_j(n)$ of the next decimation step 
according to (\ref{RG1})-(\ref{RG3}) above with $\nu=\lambda^{d-2}$. 
 
Then introducing a parameter $\alpha$, ($0\leq \alpha$), define the 
interpolating coefficients:
\beq
\tilde{c}_j(n,\alpha)= \tilde{c}_j(n-1)^{\lambda^{2}(1-\alpha)}\,
 c_j(n)^\alpha \, . 
\eeq
Then, 
\[ \tilde{c}_j(n,\alpha)= \left\{ \begin{array}{lll}
c_j(n) & : &\alpha=1 \\ 
\tilde{c}_j(n-1)^{\lambda^{2}} & : & \alpha=0 
\end{array} \right. \]
The $\alpha=0$ value is that of the $n$-th step coefficients 
resulting from (\ref{RG1})-(\ref{RG3}) with $\nu=1$.

Thus defining 

\bea    Z_\Lambda(\beta, \alpha, n) &= & 
\big(\prod_{m=0}^{n-1} F_0(m)^{|\Lambda|/\lambda^{md}}\big)\;
F_0(n)^\alpha \nonumber \\ 
& & \cdot\;
\int dU_{\Lambda^{(n)}}\;\prod_p\,f_p(U,n,\alpha) \nonumber
\eea
where 
\[f_p(U,n,\alpha)= \Big[ 1 + \sum_{j\not= 0} \tilde{c}_j(n,\alpha)
\,\chi_j(U) \Big] \, ,\]
we have from (I), (II) above: 
\[ Z_\Lambda(\beta, 0, n) \leq Z_\Lambda(\beta, n-1) 
\leq Z_\Lambda(\beta, 1, n)  \;.\]
But then, by continuity, there exist a value 
$0 < \alpha=\alpha^{(n)} < 1$ \quad  
such that 
\[ Z_\Lambda(\beta, \alpha^{(n)}, n)= Z_\Lambda(\beta, n-1) \;.\] 

So starting at original spacing $a$, at every decimation step 
$m$, ($m=0,1,\cdots,n$), there is a value $0< \alpha^{(m)}(\beta) <1$ 
such that 
\beq
Z_\Lambda(\beta, \alpha^{(m+1)}, m+1)=
Z_\Lambda(\beta, \alpha^{(m)}, m)\, .
\eeq 

This then gives an exact {\it representation} of the original 
PF in the form:
\bea    Z_\Lambda(\beta) &= & 
F_0^{|\Lambda|}\; \int dU_\Lambda\;\prod_p\,f_p(U,a) \nonumber \\
  & =& \prod_{m=0}^n F_0(m)^{\alpha^{(m)}|\Lambda|/\lambda^{md}}\;
\nonumber \\ 
& & \cdot\;
\int dU_{\Lambda^{(n)}}\;\prod_p\,f_p(U,n,\alpha^{(n)}) , \label{rep} 
\eea
i.e. in terms of the successive bulk free energy contributions from the 
$a \to \lambda \to \cdots \to \lambda^n a$ decimations 
and a single plaquette effective action on the resulting 
lattice $\Lambda^{(n)}$.

Consider now the coefficients at the, say, $n$-th step in this 
representation: \quad $\tilde{c}_j(n,\alpha=\alpha^{(n)})$. \\ 
Compare them to those evaluated at $\alpha=1$:\\ 
$\tilde{c}_j(n,\alpha=1)\equiv c_j(n)$, \ 
which will be referred to as 
the MK coefficients ($\alpha=1 \Longleftrightarrow \nu=\lambda^{d-2}$, 
the standard MK choice), and which, according to (I) give an upper bound. 

One can then prove that  
\beq
\tilde{c}_j(n,\alpha) \leq c_j(n) \qquad \mbox{for {\it any}} 
\quad 0\leq \alpha\leq 1 \;.\label{compineq}
\eeq 
This has the following important consequence. 

Assume we are in a dimension $d$ such that 
under successive decimations the MK coefficients ($\alpha=1$) 
are non-increasing. Then, (\ref{compineq}) implies: 
\bea
  \tilde{c}_j(n,\alpha^{(n)}) & \geq &  c_j(n+1) \geq  
\tilde{c}_j(n+1,\alpha^{(n+1)}) \nonumber \\
     & \geq & c_j(n+2) \geq 
\tilde{c}_j(n+2,\alpha^{(n+2)}) \nonumber \\
     &\geq & \cdots    \nonumber 
\eea

Thus, if  the $c_j(n)$'s are non-increasing, 
so are the $\tilde{c}_j(n,\alpha)$. 
The  $c_j(n)$'s must then approach a fixed point, and hence so must the 
$\tilde{c}_j(n,\alpha)$'s, since $c_j(n), \tilde{c}_j(n,\alpha)\geq 0$. 
Note the fact that this conclusion is independent of the specific 
value of the $\alpha$'s at every decimation step.

In particular, if the $c_j(n)$'s approach the strong coupling  
fixed point, i.e. $F_0\to 1$, $c_j(n) \to 0$ as $n\to \infty$, so must 
the $\tilde{c}_j(n,\alpha)$'s of the exact representation. 
{\it If the MK decimations confine, so do those in the exact 
representation (\ref{rep}).} 

As it is well-known by explicit numerical evaluation, the MK decimations 
for $SU(2)$ and $SU(3)$ indeed confine  
for all $\beta<\infty$ and $d\leq 4$.
 
Does this imply a proof of confinement for the exact theory?

Though strongly suggestive of confinement for all $\beta$ in 
the exact theory, the above does not yet constitute 
an actual proof.
The statement concerns the long distance action part in 
the representation (\ref{rep}) which also includes the 
(dominant) bulk contributions 
from integration over all scales from $a$ to $\lambda^n a$.    
The complete representation provides an equality to the 
value of the exact free energy.  This, just 
by itself, does not suffice to rigorously determine, at least in any direct way, 
the actual behavior of 
long distance order parameters characterizing phases in the exact theory. 
To do this one needs to  
carry through the above derivation for the partition function also for the 
case of the appropriate order parameters.   

Now the derivation of the basic two statements (I) and (II) above 
assumes translation invariance and reflection positivity. 
In the presence of observables 
such as a Wilson loop, translation invariance is broken and 
reflection positivity is reduced to hold only in planes bisecting the loop. 
This does not allow the above derivations to be carried through in any 
obvious way. Fortunately, there are other order parameters that 
maintain translational invariance, and are indeed the natural 
candidates in the present context since they are constructed out 
of partition functions. These are the the vortex free energy, and 
its $Z(N)$ Fourier transform (electric flux free energy). 
Recall that the vortex free energy is defined by 
\[ e^{-F_{v}} = Z_\Lambda(\tau)/Z_\Lambda \; .\] 
Here, $Z_\Lambda(\tau)$ denotes the PF with action modified by the 
`twist' $\tau \in Z(N)$ representing a discontinuous 
gauge transformation which introduces 
$\pi_1(SU(N)/Z(N))$ vortex flux winding through the 
periodic lattice in $(d-2)$ directions. The vortex free energy is then 
the ratio of the PF in the presence of this external flux to the PF in the 
absence of the flux (the latter is what was considered above). 
The above development, in particular the derivation of (\ref{rep}), should 
then be repeated also for $Z_\Lambda(\tau)$. There is a technical 
complication. The presence of the flux, though translation invariant, 
reduces reflection positivity to hold only in planes 
perpendicular to the  directions in which 
the flux winds through the lattice. Hopefully, this will suffice to 
allow a generalization of the previous derivation to go through. 
This is currently under investigation.

\end{document}